\documentclass[fleqn,10pt]{wlscirep}
\usepackage[utf8]{inputenc}
\usepackage[T1]{fontenc}
\usepackage{braket}
\newcommand{\rad}{\,\mathrm{rad}}
\newcommand{\nm}{\,\mathrm{nm}}
\newcommand{\um}{\,\textrm{µm}}  
\newcommand{\mm}{\,\mathrm{mm}}
\newcommand{\us}{\,\textrm{µs}}
\newcommand{\ba}{$^{137}\mathrm{Ba}^+$\,}
\newcommand{\ground}{$\mathrm{S_{1/2}}\,$}
\newcommand{\pone}{$\mathrm{P_{1/2}}\,$}

\newcommand{\shelf}{$\mathrm{D_{5/2}}\,$}
\title{Low Cross-Talk Optical Addressing of Trapped-Ion Qubits Using a Novel Integrated Photonic Chip}

\author[1,+,*]{A. S. Sotirova}
\author[2,+,*]{B. Sun}
\author[1]{J. D. Leppard}
\author[2]{A. Wang}
\author[2]{M. Wang}
\author[1]{A. Vazquez-Brennan}
\author[1]{D. P. Nadlinger}
\author[3]{S. Moser}
\author[3]{A. Jesacher}
\author[2]{C. He}
\author[1]{F. Pokorny}
\author[2,*]{M. J. Booth}
\author[1,*]{C. J. Ballance}
\affil[1]{University of Oxford, Department of Physics, Oxford, OX1 3PU, United Kingdom}
\affil[2]{University of Oxford, Department of Engineering Science, Oxford, OX1 3PJ, United Kingdom}
\affil[3]{Institute of Biomedical Physics, Medical University of Innsbruck, Müllerstraße 44, 6020 Innsbruck, Austria}

\affil[*]{Corresponding to: B. Sun (b.s.shawnsuen@gmail.com), or A. S. Sotirova (ana.sotirova@physics.ox.ac.uk), or M. J. Booth (martin.booth@eng.ox.ac.uk), or C. J. Ballance (chris.ballance@physics.ox.ac.uk) }

\affil[+]{These two authors contributed equally to this work.}

\begin{abstract}
Individual optical addressing in chains of trapped atomic ions requires generation of many small, closely spaced beams with low cross-talk. Furthermore, implementing parallel operations necessitates phase, frequency, and amplitude control of each individual beam. Here we present a scalable method for achieving all of these capabilities using a novel integrated photonic chip coupled to a network of optical fibre components. The chip design results in very low cross-talk between neighbouring channels even at the micrometre-scale spacing by implementing a very high refractive index contrast between the channel core and cladding. Furthermore, the photonic chip manufacturing procedure is highly flexible, allowing for the creation of devices with an arbitrary number of channels as well as non-uniform channel spacing at the chip output. We present the system used to integrate the chip within our ion trap apparatus and characterise the performance of the full individual addressing setup using a single trapped ion as a light-field sensor. Our measurements showed intensity cross-talk below $10^{-3}$ across the chip, with minimum observed cross-talk as low as $\mathcal{O}\left(10^{-5}\right)$. 

\end{abstract}
\begin{document}
\flushbottom
\maketitle

\thispagestyle{empty}

\section*{Introduction}
\label{sec:intro}
Since their original proposal as a platform for quantum information processing\cite{cirac_quantum_1995}, trapped ions have emerged as one of the leading contenders for building a useful quantum computer. To date, small-scale trapped-ion systems have demonstrated the highest single- and two-qubit gate fidelities\cite{harty_high-fidelity_2014,ballance_high-fidelity_2016,clark_high-fidelity_2021, srinivas_high-fidelity_2021}, longest coherence times\cite{wang_single_2021}, and lowest state preparation and measurement errors\cite{an_high_2022} of any quantum computing platform. For realising a functional large-scale quantum computer, this level of control needs to be extended to a large number of qubits. This includes the implementation of all necessary quantum operations on the qubit register with very high fidelity as well as the execution of targeted (addressed) operations on specific subsets of qubits within the register with minimal effect on the unused (idle) qubits in the computation.

\begin{figure}[!ht]
    \centering
    \includegraphics
    {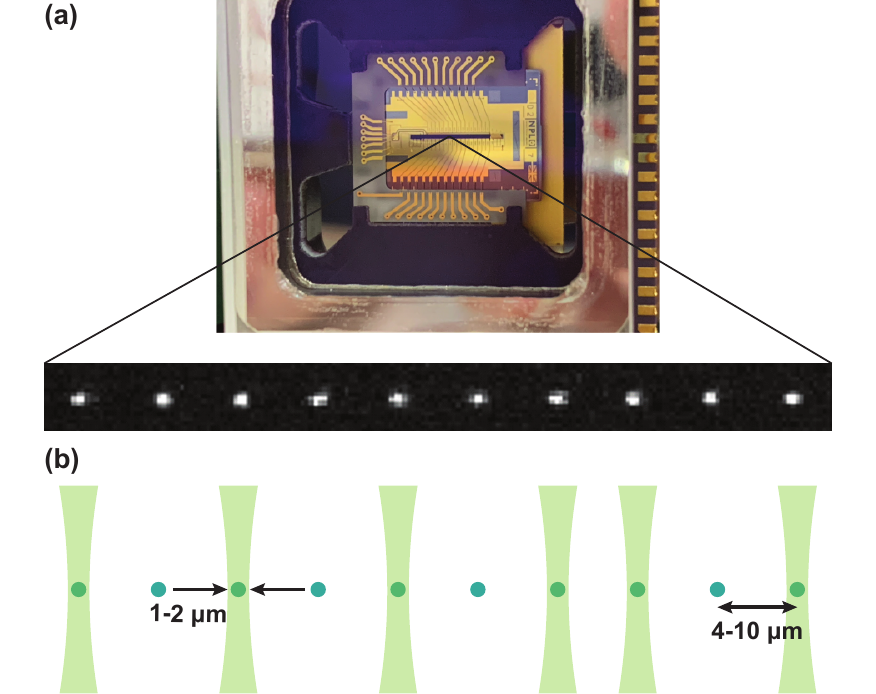}
    \caption{Individual optical addressing requirements in typical ion trap experiments. (a) A linear chain of \ba ions confined in a 3D segmented trap\cite{wilpers_compact_2013, choonee_silicon_2017}. The ions in this image are uniformly spaced with a mean ion-ion separation of $9.6(3)\um$. The image was taken using an sCMOS camera detecting light scattered from the ions. (b) Requirements for individual optical addressing in chains of trapped ions. The spacing between the individual laser beams must match the ion separation, typically $4-10\um$\cite{wineland_experimental_1998, nagerl_ion_1998}. The spot size needs to be small enough to minimise the intensity at the neighbouring ions, but larger than the diffraction limit for the laser wavelength in use.}
    \label{fig:individual_addressing_concept}
\end{figure}

Individual addressing in most types of ion trap quantum computing architectures requires locally modifying the interaction between the ions and the radiation used to perform operations with very low cross-talk, where the natural inter-ion spacing is a few micrometres\cite{wineland_experimental_1998, nagerl_ion_1998}. This can be done by altering the properties of the magnetic and/or electric fields experienced by each of the ions to change their response to a globally applied field\cite{leu_fast_2023, srinivas_coherent_2023, leibfried_individual_1999, warring_individual-ion_2013, wang_individual_2009, seck_single-ion_2020}, by focusing down the radiation used to perform quantum gates on each ion\cite{crain_individual_2014, shih_reprogrammable_2021, wang_high-fidelity_2020, pogorelov_compact_2021, egan_scaling_2021, binai-motlagh_guided_2023, timpu_laser-written_2022} as shown in fig.~\ref{fig:individual_addressing_concept}, or by using a combination of the two approaches. The former approach works with both laser and microwave radiation but requires intricate local control of electric and magnetic fields. Hence it is only compatible with microfabricated traps with numerous electrodes. The latter approach is only viable when optical wavelengths are used for operations, either when the qubit transition is driven with a two-photon Raman process or the qubit transition itself corresponds to an optical wavelength. It requires focusing down laser beams to micrometre scales, therefore working close to the diffraction limit. However, this method is applicable to all ion trap architectures. 

Several methods have been developed for achieving individual optical addressing in ion trap systems, including micro-mirror beam steering\cite{crain_individual_2014, shih_reprogrammable_2021, wang_high-fidelity_2020}, acousto-optic deflectors\cite{pogorelov_compact_2021}, multi-channel acousto-optic modulators\cite{egan_scaling_2021} (AOMs), and micro-lens arrays\cite{binai-motlagh_guided_2023}. These methods vary in cross-talk performance and scalability. Using micro-mirror arrays to steer the lasers onto the target ions offers individual addressing with very low cross-talk. However, the time required to reconfigure the beam positions or to tune the amplitude of the beams is comparable to the timescale of the gate operations, therefore accounting for a significant amount of the sequence run time. Acousto-optic deflectors offer similarly low cross-talk but also enable fast intensity control of the individual beams. However, they lack individual beam frequency control, therefore limiting the set of unitaries that can be implemented in parallel. Furthermore, only a few ions can be addressed at any one time due to power losses in higher diffraction orders. Multi-channel AOMs solve the problem of fast parallel control. However, they exhibit an order of magnitude larger cross-talk compared to the previous two approaches due to electronic cross-talk between channels in existing devices. Additionally, they are only manufactured with a fixed and equal inter-channel spacing, and readily available with only up to 32 channels, limiting both scalability and ability to address non-evenly spaced chains. The micro-lens array approach solves the majority of the problems listed above by enabling parallel operations with very low cross-talk. However, the system presented in ref.~\citenum{binai-motlagh_guided_2023} has a very high insertion loss limiting the obtainable light intensity at the ion, and a fixed, uniformly spaced output pattern that cannot be easily adapted to non-uniform ion crystals.

In this work, we demonstrate a novel approach to individual optical addressing in trapped-ion chains with minimal cross-talk using a network of fibre-coupled modulators connected to a high-performance photonic chip. We employ spherical phase-induced multiscan waveguides\cite{sun_-chip_2022} (SPIM-WGs) that provide precise control over the optical mode and enable much higher refractive index (RI) contrast modifications in optical glass compared to conventional ultrafast laser-written waveguides. A similar approach using laser-written waveguide devices has been reported\cite{timpu_laser-written_2022}, however 
matching the output of these waveguides to the ion chain while maintaining a good spot quality and low cross-talk is yet to be demonstrated.

The photonic chip that we present here adopts individual adiabatic mode converters as light guiding channels that exhibit excellent optical mode confinement and low cross-talk even at a channel separation of a few micrometres. This ensures that the errors due to nearest-neighbour cross-talk do not limit the performance of the trapped-ion device. Furthermore, the manufacturing process of the SPIM-WGs offers high flexibility, facilitating easy modification of the number of channels, the channel positions, and the mode shapes in the chip design, therefore making the chip easily adaptable to the ion configurations found in most ion trap experiments. The use of a fibre network for light delivery to the photonic chip further simplifies the process of exchanging devices by minimising the necessary realignment, thus allowing for rapid iteration of the system design.

\section*{Results}
\label{sec:results}

\subsection*{Photonic chip design}
\label{sec:chip-design}
Performing laser-driven targeted operations in long chains of trapped ions imposes several competing requirements on the individual addressing setup. First, to minimise errors on the idle qubits, it is crucial to minimise the cross-coupling between neighbouring ion sites. Second, the spacing between neighbouring channels must match the ion-ion spacing, typically on the order of a few micrometres \cite{wineland_experimental_1998, nagerl_ion_1998}. This requires generating a series of closely spaced beams, each focused to an $\mathcal{O}\left(1\um\right)$ waist radius as shown in fig.~\ref{fig:individual_addressing_concept}(b). As a result, this approach necessitates focusing the beams near the diffraction limit. Working with tightly focused beams can also increase errors in the quantum operations due to intensity, phase, and/or polarisation modulation of the light at the ion position. This modulation can be caused by mechanical drifts in the optical system or by the secular motion of the ions\cite{cetina_control_2022, west_tunable_2021}. Hence it is desirable to make the beam waist radius as large as possible. The requirements on the beam spacing and beam waist radius constrain the waist-to-spacing ratio of the device output. A larger ratio makes integration into the trapped-ion system easier and more robust to errors but also increases the cross-talk within the device. Furthermore, to ensure that the required input laser power scales favourably with the qubit register size, it is important to maintain a high throughput efficiency of the system. Finally, the ability to control the intensity, phase, and frequency of individual channels in parallel enables the simultaneous application of different unitary operations on different target qubits, therefore reducing the algorithm run time.

\begin{figure}[!ht]
    \centering
    \includegraphics{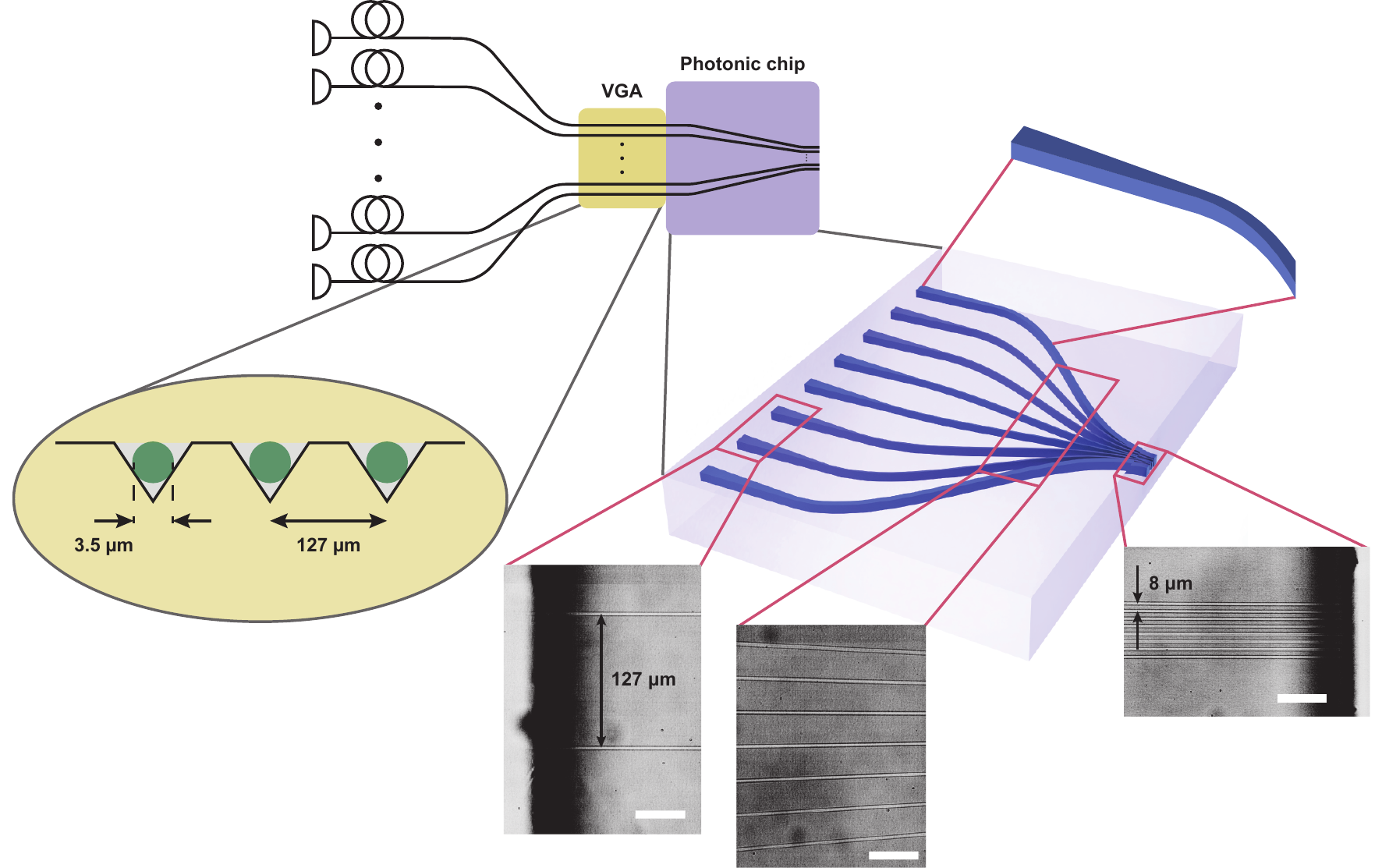}
    \caption{A diagram of the photonic chip design. The input light from the V-groove array (VGA), whose output channels are spaced by $127\um$, is coupled into the photonic chip as shown in the top left. Shown in the middle is a 3D representation of a photonic chip featuring eight channels. Each channel serves as a high-efficiency adiabatic mode converter. The optical modes are precisely engineered to satisfy the trapped-ion addressing requirements at the output of the photonic chip, where the channel separation is brought down to $8\um$. The grayscale microscopic images of the channel structures were acquired from the top of the photonic chip. The scale bar is $50\um$. The dark regions at the chip facets were due to reduced microscopic illumination. The device has two straight regions, input ($2.2\mm$) and output ($0.2\mm$), which are connected by a curved region ($7.2\mm$). The adiabatic mode conversion is performed over the $2.2\mm$ at the channel input and the corresponding region is highlighted at top right of the diagram.}
    \label{fig:v-groove-waveguide}
\end{figure}

To address these challenges, we developed a novel photonic chip connected to a network of optical fibre components. In our system the source laser is coupled into a series of fibre splitters, such that the light is split up into the required number of channels. Each channel is then connected to a fibre AOM that allows for individual phase, frequency, and amplitude control, and for switching of each beam. The fibre AOMs are connected to a fibre V-groove array (VGA) whose output is an array of fibre cores with a mode field diameter (MFD) of $3.5\um$ and a uniform spacing of $127\um$. The VGA is then coupled to the photonic chip as shown in fig.~\ref{fig:v-groove-waveguide} and fixed in place with glue (details in "Materials and methods") to avoid misalignment during operation. 

The photonic chip performance benefited from several advanced design and fabrication techniques, as detailed in the following sections. It was designed to convert the input optical modes and spacing of the VGA channels into closely spaced ($8\um$) modes suitable for trapped-ion addressing. Each channel incorporated a high-efficiency adiabatic mode converter to transform the optical mode from the VGA, equivalent to a single-mode fibre (SMF), into the smaller optical mode required to address the ions. The adiabatic mode conversion was implemented over a distance of $2.2\mm$, starting from the chip's input facet. Subsequently, curved routing of the channels was used to reduce the channel spacing, while maintaining their cross-sectional shape at the output facet of the chip. The microscopic images in fig.~\ref{fig:v-groove-waveguide} obtained from the top of the photonic chip, capturing the input facet, bending region, and output facet, illustrate the progression of the channel spacing along the chip. Even though the channels were closely stacked with $8\um$ spacing over a distance of $0.2\mm$, we were able to maintain a low nearest-neighbour cross-talk at a level of $\mathcal{O}\left(10^{-4}\right)$ as demonstrated in the following sections.

\subsection*{High-contrast refractive index modification}
\label{sec:high-RI-mode-degisn}
As outlined in the previous section, individual optical addressing of trapped ions requires the generation of closely spaced beams with maximum waist-to-spacing ratio and minimum cross-talk. That way, the errors due to unwanted operations on the idle qubits are minimised, while maximising the robustness of the system. This translates to the need for a very high level of light confinement in the channels of the photonic chip. Existing methods of producing micro-waveguides in photonic chips require a very small waist-to-spacing ratio to satisfy the low cross-talk requirement. An ideal fabrication method would allow an increase in waveguide size while maintaining single-mode operation and low cross-talk. The key to this is a stronger confinement of the mode by increasing the RI contrast between the core and the cladding. This level of RI control has not yet been possible with conventional ultrafast laser writing, where high RI contrast is typically accompanied by poor control of the mode shape.

\begin{figure}[!ht]
    \centering
    \includegraphics{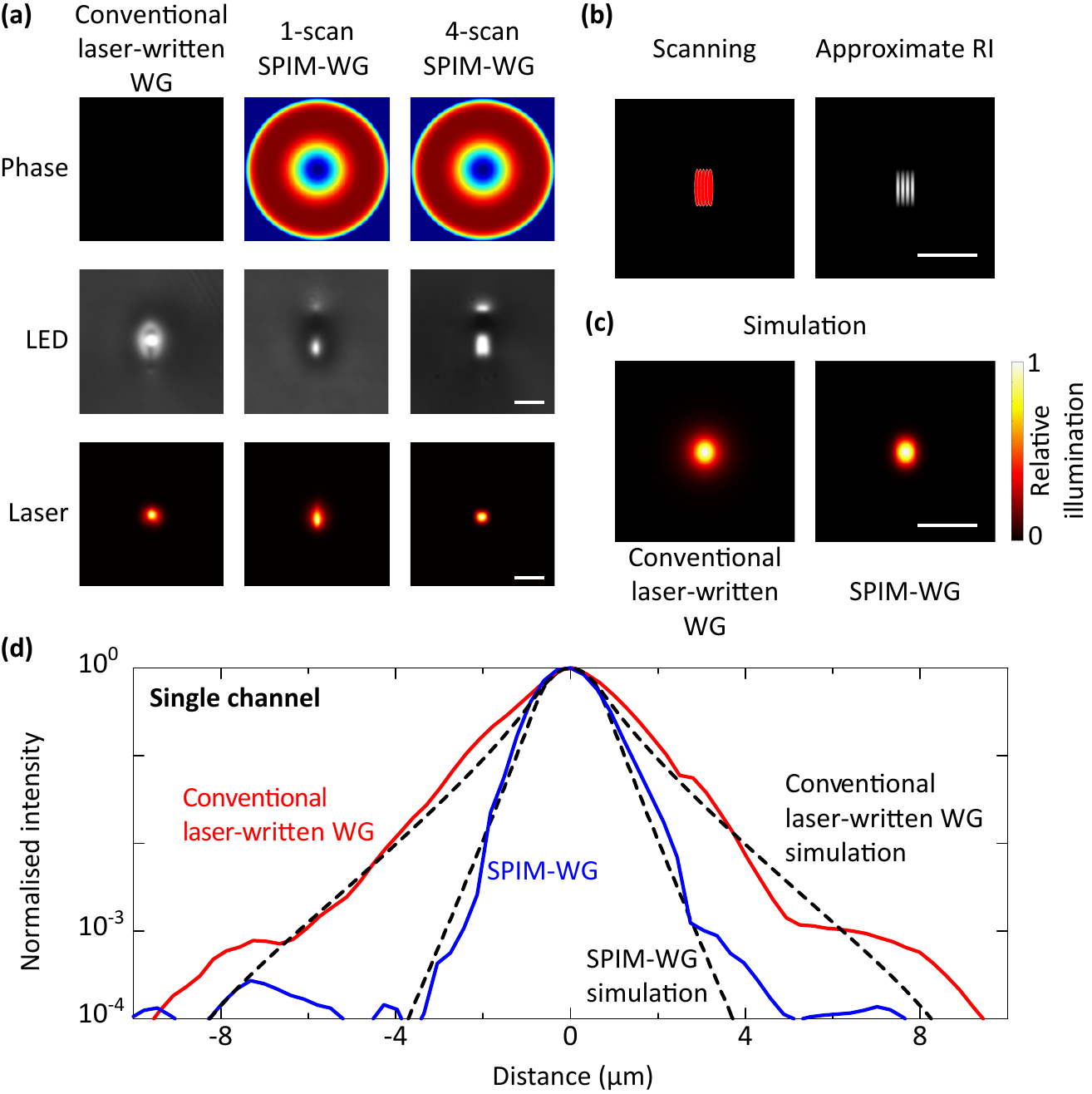}
    \caption{Optical modes at the photonic chip output. (a) Comparison of waveguide fabrication techniques: conventional ultrafast laser-written waveguide (left), single-scan SPIM waveguide (middle), and multiscan SPIM waveguide (right). Top: varying optical phases applied to the ultrafast laser system for waveguide inscription. Middle: microscope images of the waveguide facets at the chip output under broadband LED illumination. Bottom: $532\nm$ laser mode profiles of the waveguides at the chip output (intensity normalised individually). (b) Scanning scheme and approximate RI profile for a single SPIM-WG channel at the chip output. (c) COMSOL-simulated mode profiles for a conventional laser-written waveguide and for the designed SPIM-WG output (intensity normalised individually). (d) HDRMs and COMSOL-simulated optical modes for a conventional laser-written waveguide and for the designed SPIM-WG. }
    \label{fig:high_RI_contrast}
\end{figure}

In order to address this challenge, in this work we fabricated SPIM-WGs using a multiscan scheme to precisely control the shape and the size of the channel cross-section, enabling single-mode operation and a large core diameter. To achieve high-precision RI modification, the SPIM-WGs were fabricated with combined spherical aberrations of first-order Zernike mode $11$ and third-order Zernike mode $37$ introduced into the focused laser\cite{sun_high_2023}, with root mean square (RMS) amplitudes of $-1\rad$ and $-0.3\rad$, respectively. The SPIM-WG design incorporated four scans with a core separation of $0.4\um$ forming a single light guiding channel. The RI contrast measured using high-resolution quantitative phase microscopy was approximately $0.015$ (with a core RI of $1.525$ and a cladding RI of $1.51$), which is two to three times higher than for a waveguide created by conventional ultrafast laser writing\cite{sun_-chip_2022}. The horizontal core size of one complete SPIM channel was measured to be about $1.8\um$, which was approximately the maximum diameter that maintained single-mode operation with an RI contrast of $0.015$ at $532\nm$\cite{snyder_optical_1984}. Figure \ref{fig:high_RI_contrast}(a) presents a comparison between LED microscopic images and $532\nm$ laser mode profiles for a conventional laser-written waveguide, a single-scan SPIM-WG, and a four-scan SPIM-WG. Figure \ref{fig:high_RI_contrast}(b) illustrates the scanning scheme and the approximate RI profiles. Additionally, fig.~\ref{fig:high_RI_contrast}(c) shows COMSOL-simulated mode profiles for a conventional laser-written waveguide and a four-scan SPIM-WG. We observed a slight mode elongation along the vertical direction. However, this does not affect the mode quality or the cross-talk in the direction parallel to the ion chain. In fact, it is advantageous for optical addressing of trapped ions, as it improves the robustness of the system to intensity modulation in that direction.

We observed a bright lobe above the four-scan SPIM-WG as shown in fig.~\ref{fig:high_RI_contrast}(a). Such lobes were significantly weaker in the single-scan SPIM-WG. Close examination showed that these lobes possessed a low RI contrast and exhibited high guiding losses. Notably, there was a negative relative RI modification between the main lobe and the upper lobe. In this application, the main lobe was used as the light guiding channel, and we observed no adverse effects arising from the presence of the upper lobe. The strong confinement to the main guiding channel could be attributed to the presence of the region with negative RI contrast between the two lobes, as well as the high transmission loss associated with the upper lobe.

To illustrate the light confinement capabilities, we performed high dynamic range measurements (HDRMs) of the guided laser modes at $532\nm$. Figure \ref{fig:high_RI_contrast}(d) illustrates measured and simulated HDRMs for a single channel. We observed a close agreement between the simulations and the experiments. A clear difference in the degree of mode confinement was observed between the conventional laser-written waveguide and the SPIM-WG. At distances of $2-8\um$ from the centre, the intensity from the SPIM-WG was around an order of magnitude lower compared to that of the conventional laser-written waveguides we tested.

\begin{figure}[!ht]
    \centering
    \includegraphics
    {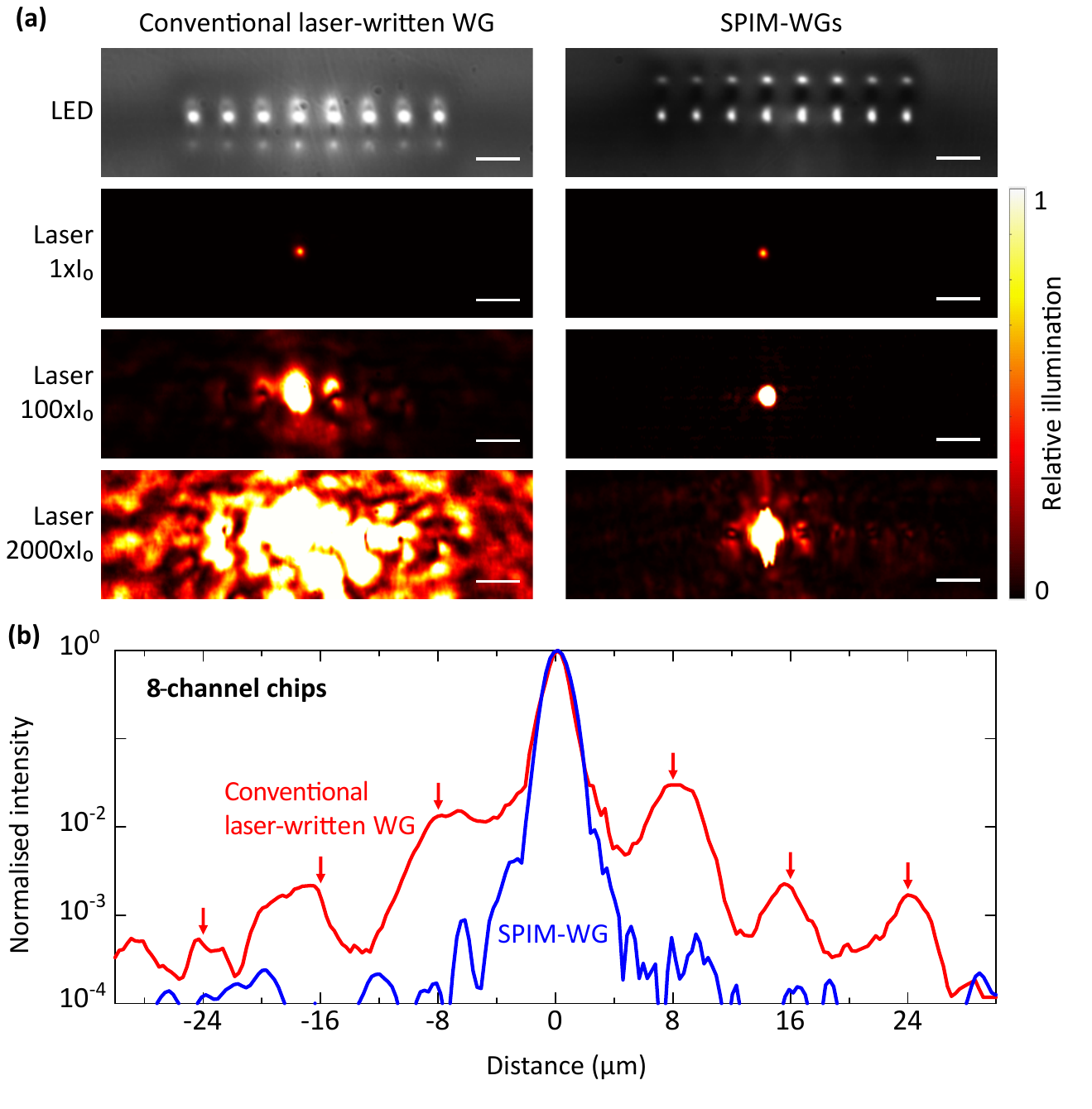}
    \caption{Light confinement properties of the SPIM-WG channels. (a) LED-illuminated microscope images and laser mode profiles for two 8-channel chips fabricated using conventional laser-written (left) and the SPIM-WG (right) methods. One SMF was coupled to the fourth channel (left to right) of the chip's input facet. To show effects over a high dynamic range, three laser intensities ($1\times I_0$, $100\times I_0$, and $2000\times I_0$, adjusted using neutral density filters) were applied. (b) HDRMs for two 8-channel chips, each fabricated using conventional laser-written or the SPIM-WG methods. The red arrows mark the positions of the remaining channels.}
    \label{fig:convWG_vs_SPIM}
\end{figure}

Eight-channel chips were designed according to the specifications outlined in the previous section and illustrated in fig.~\ref{fig:v-groove-waveguide}. In fig.~\ref{fig:convWG_vs_SPIM}(a) we show LED microscopic images of the output chip facet. The outer channels appear dimmer than the central channels due to the lower LED intensity away from the centre. We conducted measurements to assess the overall loss (including coupling and propagation losses) across all waveguide channels and found the bending losses in all channels to be negligible, owing to the high RI contrast resulting in a strong mode confinement. 

Channel cross-talk was assessed by coupling a $532\nm$ laser through a single-mode fibre to one central channel at the chip input facet, then observing the full 8-channel intensity distribution at the chip output. High dynamic range measurements were compiled from a sequence of images taken with different calibrated neutral density filters. As shown in fig.~\ref{fig:convWG_vs_SPIM}(a), the SPIM-WGs showed far better confinement of the laser light to the vicinity of the channel compared to conventional laser-written waveguides. The light intensity at the position of the neighbouring channels, i.e. $8\um$ away from the channel, was one to two orders of magnitude lower for the SPIM-WGs. Further details are provided in fig.~\ref{fig:convWG_vs_SPIM}(b). The red curve, corresponding to conventional laser-written waveguides, clearly shows coupling into the neighbouring channels, evident from the multiple high light intensity peaks at $8\um$ intervals corresponding to the channel separation. In contrast, the SPIM-WGs exhibited minimal coupling to adjacent channels. We measured nearest-neighbour cross-talk of $\approx 3\times10^{-2}$ for the conventional laser-written waveguides and $\approx 5\times10^{-4}$ for the SPIM-WG channels. These measurements were repeated for multiple different chips, all of which showed consistent results, confirming the high reliability of the chip fabrication.

\subsection*{Advanced mode matching and adiabatic mode conversion}
\label{sec:mode-matching}
\begin{figure}[!ht]
    \centering
    \includegraphics{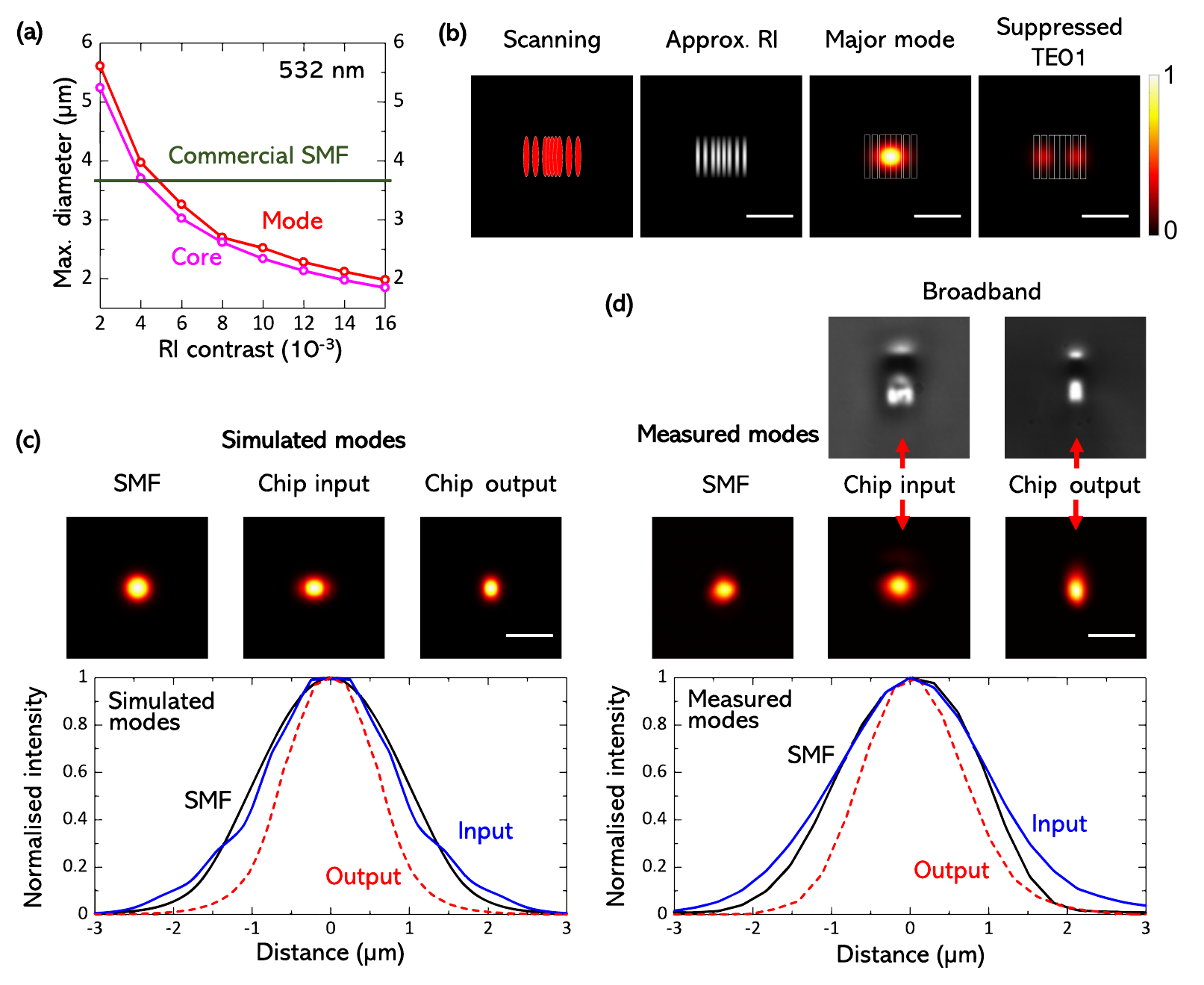}
    \caption{Enhancing chip coupling efficiency through advanced mode matching and adiabatic mode conversion. (a) Calculated maximum waveguide core diameter to maintain single-mode guiding at $532\nm$ as a function of the RI contrast (core RI minus cladding RI). The maximum MFDs were determined from a COMSOL simulation. The cladding RI was $1.51$ for borosilicate Eagle glass at $532\nm$. Commercial SMFs typically have a cladding RI of 1.455 and a core RI of $1.4607$. (b) Scanning scheme, approximate RI profile, and COMSOL-simulated guiding modes for a single SPIM-WG channel at the chip's input facet. (c) Top: COMSOL-simulated laser guiding modes for an SMF, the output channel, and the input channel of the chip. Bottom: the dominant mode of the chip's input channel compared to the SMF mode and the chip's output mode. (d) Top: LED-illuminated microscope images of the chip's output and input channels. Middle: experimentally measured $532\nm$ laser guiding modes for an SMF and for the SPIM channels. Bottom: intensity plots of the measured modes.}
    \label{fig:AMM_conversion}
\end{figure}

To ensure that the required laser input power to the addressing setup scales favourably with the number of qubits, it is crucial to maintain a high throughput efficiency within the chip-based system. This has been a problem in previous single-ion addressing implementations, where losses of more than $10\,\mathrm{dB}$ have been observed due to poor coupling between system components\cite{binai-motlagh_guided_2023}. The photonic chip must therefore implement high-efficiency coupling between the set of SMFs delivering the light, held in a VGA, to the input of the ion-trap lens system. This can only be achieved through effective mode matching between the components and adiabatic conversion between the modes. Figure \ref{fig:AMM_conversion}(a) shows calculated maximum diameters for the waveguide core and guiding mode to maintain single-mode operation for different RI contrasts. Commercial SMFs typically have an MFD of $3.5-3.7\um$ at a wavelength of $532\nm$, while the SPIM-WG channel has a maximum single-mode MFD of $1.9\um$ along the horizontal direction, owing to the high RI contrast of $0.015$. Significant coupling losses would arise if the larger SMF modes are coupled directly to the smaller, high RI SPIM-WG channel modes.
 
An additional concern relates to position uncertainties in the VGA. The commercial VGAs we used were designed to have a core separation of $127\um$, but exhibited variable position offsets between the fibre cores of up to $0.7\um$ along the $x$-direction (parallel to the linear fibre core array) and up to $0.3\um$ along the $y$-direction (orthogonal to the linear fibre core array). These offsets significantly reduced the mode overlaps between the VGA and the photonic chip and increased the coupling losses. 

We developed a novel design for advanced mode matching that effectively mitigates the impact of the VGA channel position variability. The design, presented in fig.~\ref{fig:AMM_conversion}(b), was built upon the central region of the four scans discussed earlier [fig.~\ref{fig:high_RI_contrast}(b)]. We introduced two additional scans on either side of the 4-scan SPIM-WG in order to increase the lateral size of the waveguide mode. The spacing of these additional scans was $1.5$ times larger than that of the four central scans. As shown in fig.~\ref{fig:AMM_conversion}(b), this design considerably extended the lateral size of the dominant mode while effectively suppressing higher-order modes (e.g., the TE01 mode). Figure \ref{fig:AMM_conversion}(d) includes broadband-light-illuminated microscope images of the channel input and output cross-sections, further highlighting the difference between RI modifications.

Figure \ref{fig:AMM_conversion}(c) presents the simulated guiding modes for the SMF, the output channel (as described in the previous section), and the input channel (dominant mode only). The input channel mode was nearly identical to that of the SMF, but considerably larger than that of the output channel. The experimentally measured mode profiles, summarized in fig.~\ref{fig:AMM_conversion}(d), agreed well with the simulations. The measured mode for the input channel in fig.~\ref{fig:AMM_conversion}(d) appeared slightly larger than the simulated dominant mode in fig.~\ref{fig:AMM_conversion}(c) because the measured mode profile contained a superposition of multiple modes, while the simulation included only the dominant mode. Loss measurements indicated that this specialized design enhanced the coupling efficiency from less than $60\%$ to approximately $80\%$ and significantly improved the mode uniformity across the channels.

To address the disparity in cross-section between the channel input mode and the required output mode, we incorporated adiabatic mode conversion by changing the waveguide properties along the chip. Starting from the input facet, the RI profile was gradually changed over a total length of $2.2\mm$, transitioning from the design illustrated in fig.~\ref{fig:AMM_conversion}(b) to the design described in fig.~\ref{fig:high_RI_contrast}(b). The RI profile from fig.~\ref{fig:high_RI_contrast}(b) was then maintained throughout the bending region until the chip output. A 3D render of the adiabatic mode conversion is presented in the "Materials and methods" section. The mode conversion efficiency was investigated through measurements of the losses of straight SPIM-WG channels with and without adiabatic mode conversion. Negligible differences in loss were found between the two designs, proving the practical effectiveness of adiabatic mode conversion using SPIM-WGs.

\subsection*{Integration with a trapped-ion quantum system}
\label{sec:ion-trap-results}
In our experiment we use trapped \ba ions as qubits. The ions are confined in a 3D monolithic microfabricated trap\cite{wilpers_compact_2013, choonee_silicon_2017} that allows for generation of deep confining potentials while maintaining low heating rates, which is crucial for storing long ion crystals. The segmented electrode structure provides sufficient degrees of freedom for both ion shuttling across the trap and generation of anharmonic potential shapes. The latter is particularly important for maintaining a uniform ion-ion spacing in large registers\cite{lin_large-scale_2009}, therefore increasing the minimum distance between neighbouring ions compared to harmonically spaced chains\cite{wineland_experimental_1998}. As explained in the previous sections, a larger ion-ion spacing reduces the cross-talk between neighbouring channels in the addressing setup.

\begin{figure}[!ht]
    \centering
    \includegraphics{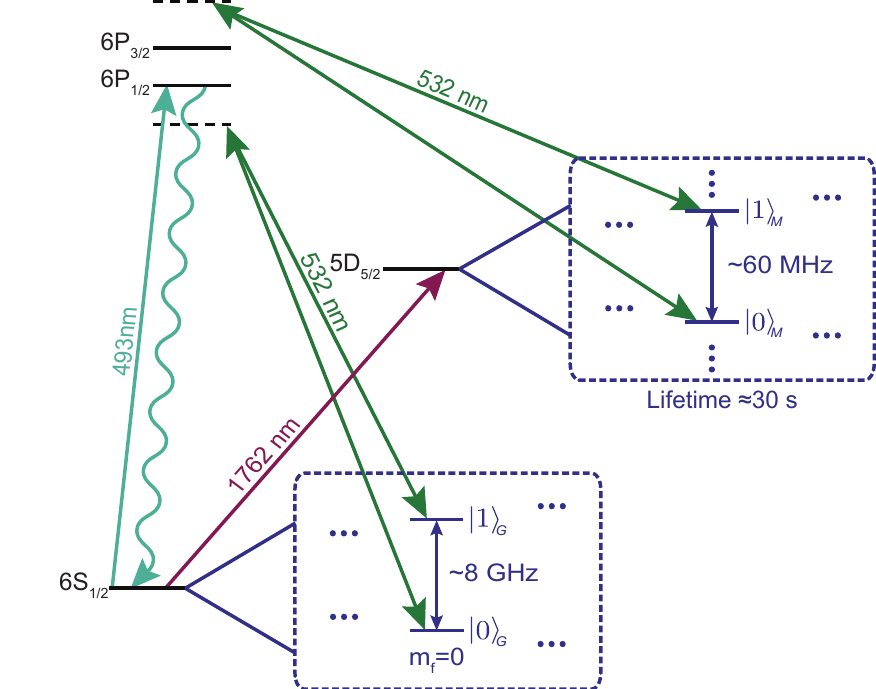}
    \caption{Level structure of \ba. Both the ground \ground and the metastable \shelf level qubit transitions can be driven with a two-photon Raman process using $532\nm$ light. The lifetime of the metastable level is $30.14\,\mathrm{s}$\cite{zhang_branching_2020}, much longer than the typical timescale of the operations, making it a suitable place to encode qubits in addition to the ground level. The ions are detected by repeatedly exciting the transition between the \ground and the \pone levels and collecting the scattered $493\nm$ light.}
    \label{fig:ba-levels}
\end{figure}

The SPIM-WGs used for individual addressing in our setup were optimised for $532\nm$. This wavelength enables driving Raman transitions within both the ground and the metastable level manifolds of the \ba ions as shown in fig.~\ref{fig:ba-levels}. The output of the photonic chip is mapped on the ion chain using a 2:1 lens relay, such that the distance between neighbouring beams at the ion position is reduced to $4\um$, matching our target ion spacing.

\begin{figure}[!ht]
    \centering
    \includegraphics{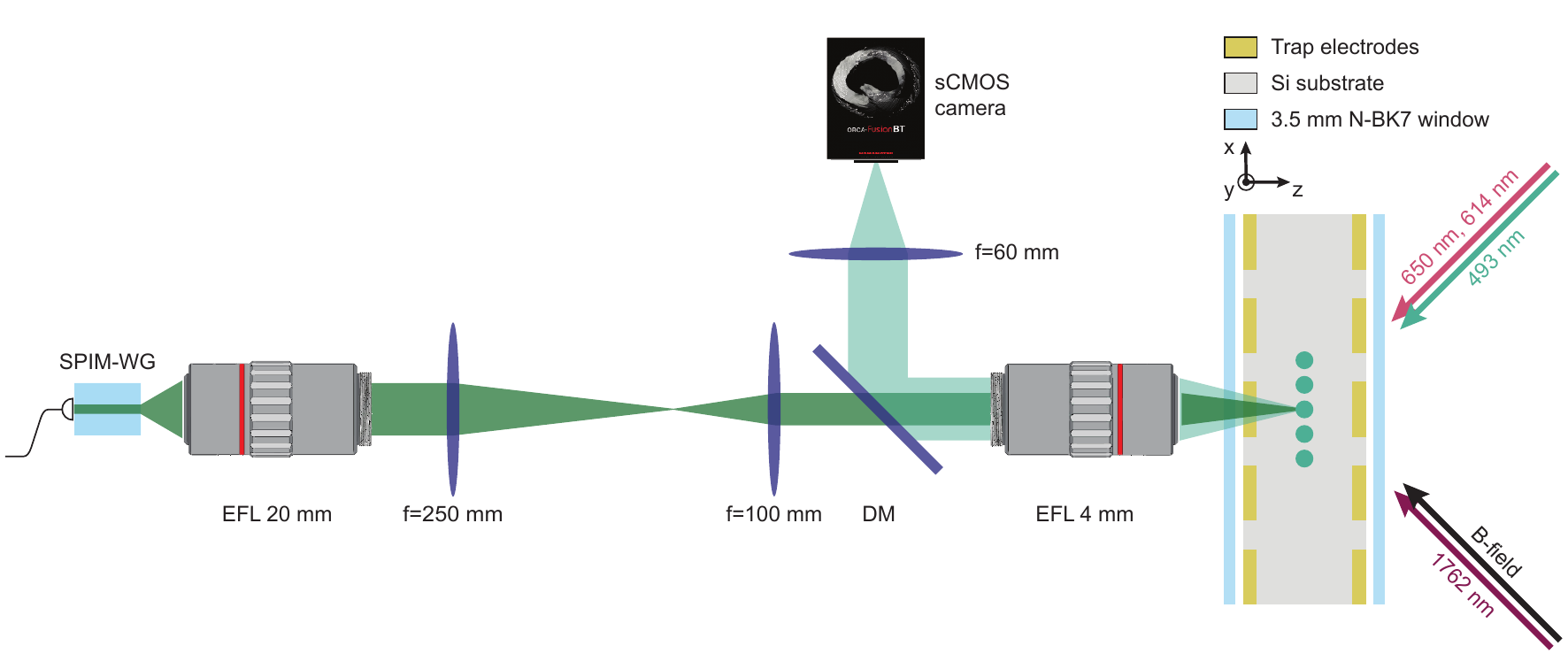}
    \caption{An overview of the trapped-ion setup. The ions are confined in a 3D monolithic microfabricated trap\cite{wilpers_compact_2013, choonee_silicon_2017}. The $493\nm$ light scattered by the ions is collimated using a commercial $\mathrm{NA}0.5$ microscope objective with an effective focal length (EFL) of $4\mm$ and refocused on an sCMOS camera for spatially resolved readout. The $532\nm$ light from the photonic chip is first collimated using a commercial microscope objective with an EFL of $20\mm$. The focal lengths and positions of the subsequent $f=250\mm$ and $f=100\mm$ lenses are chosen to achieve the required 2:1 demagnification while also satisfying the geometrical constraints of the optical system. A dichroic mirror (DM) is used to overlap the $532\nm$ beam path with the $493\nm$ fluorescence beam path. This allows us to focus the $532\nm$ light on the ions with the same NA0.5 objective we use for imaging them.}
    \label{fig:ion-trap-setup}
\end{figure}

An outline of the optical system used to reimage the SPIM-WG output on the ions is shown in fig.~\ref{fig:ion-trap-setup}. The output of the chip is collimated using a high-quality commercial microscope objective to minimise aberrations. Another microscope objective is used to both refocus the $532\nm$ control light from the waveguide on the ions, and to collect the $493\nm$ photons scattered by the ions. The latter objective is chromatically corrected for visible wavelengths, and glass-thickness compensated for the glass window on the vacuum chamber. This allows us to both image and address the ions with minimal aberrations.

The VGA and SPIM-WG assembly is mounted on a stainless steel plate before integration with the rest of the optical system (see "Materials and methods"). This, combined with the fibre network used to interface between the laser source and the photonic chip as outlined in fig.~\ref{fig:v-groove-waveguide}, enables simple exchange of chips in the ion trap system with minimal realignment.

\subsection*{Cross-talk measurement using a single trapped ion}
To characterise the performance of the fully integrated individual addressing setup, we measured the beams' spatial profiles by using a single ion as a point-like sensor. When a single $532\nm$ beam is directed at the ion, it introduces an AC Stark shift on the quadrupole transition frequencies between states in the \ground and \shelf levels. This shift is proportional to the intensity of the $532\nm$ light experienced by the ion. To characterise the spatial intensity distribution from the SPIM-WG output, we transported the ion from the trap centre to an axially displaced position $x$ (see fig.~\ref{fig:ion-trap-setup}) and measured the quadrupole frequency shift as a function of $x$. We were thus able to measure the beam profiles, the beam spacing, and the cross-talk of the system.

For the purpose of this measurement we used the $\ket{S_{1/2}, F=2, m_F=0}\leftrightarrow\ket{D_{5/2}, F=4, m_F=+1}$ transition as it exhibits the lowest magnetic field sensitivity of all available transitions for the magnetic field direction and the beam geometry shown in fig.~\ref{fig:ion-trap-setup}. The frequency shift introduced by the $532\nm$ beam can be seen as a $Z_\phi$ rotation on the qubit state with the phase $\phi$ proportional to the magnitude of the AC Stark shift, and hence proportional to the beam intensity. To estimate this phase in a way that is robust to additive errors, such as those accumulated during state preparation and measurement, we used a robust phase estimation protocol\cite{kimmel_robust_2015, rudinger_experimental_2017} (RPE). To enhance the system's coherence time and hence increase the available probe duration (therefore also increasing the dynamic range of the measurement), we embedded the RPE sequence inside a Knill dynamical decoupling sequence\cite{souza_robust_2011} (KDD). Further details on the sequences used can be found in the "Materials and methods" section.

\begin{figure}[!ht]
    \centering
    \includegraphics{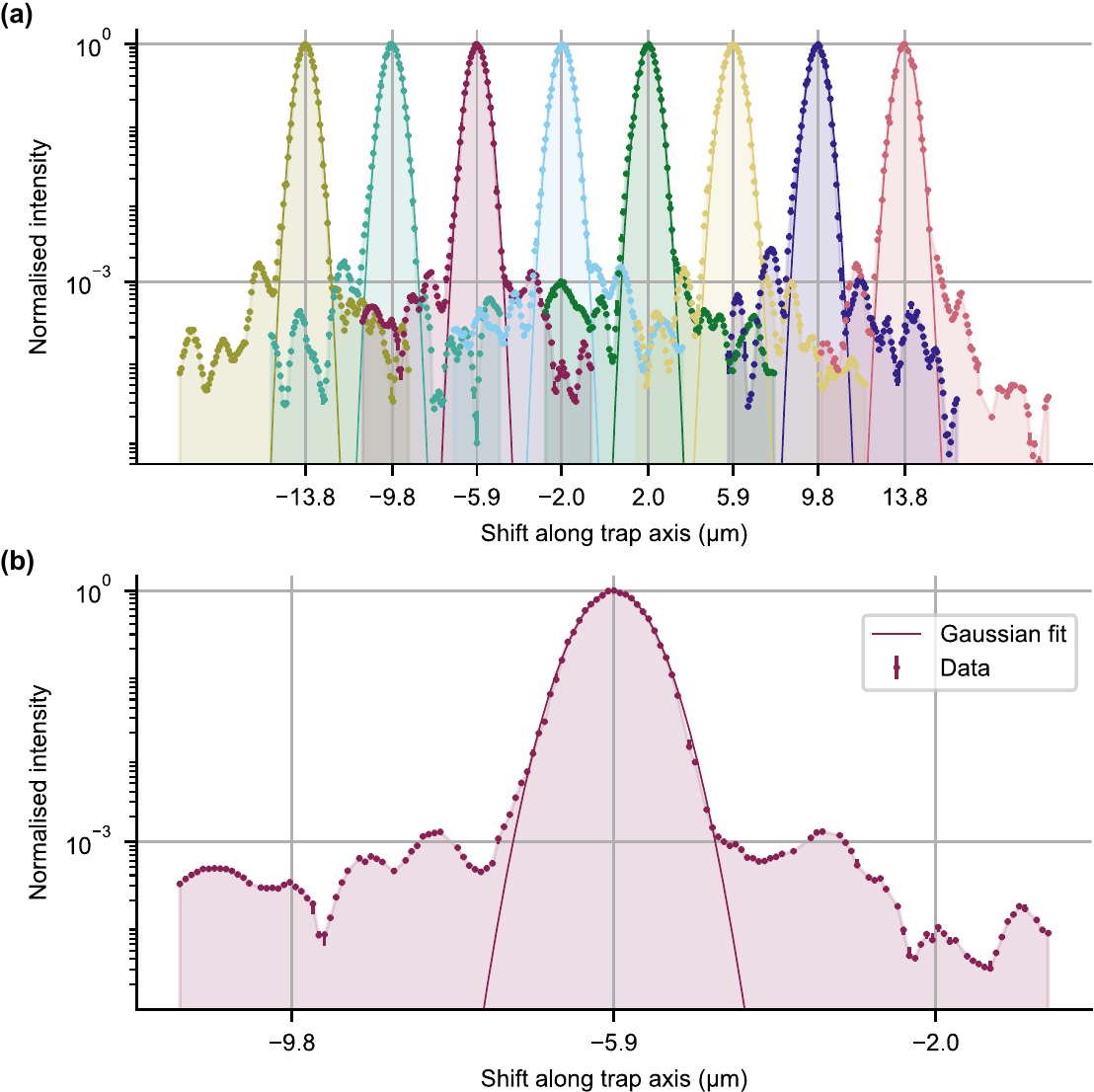}
    \caption{Intensity measurement of the $532\nm$ beams at the trap centre obtained by shuttling a single ion along the trap axis and measuring the AC Stark shift introduced on one of the quadrupole transitions between the \ground and \shelf levels. The data from each of the channels is normalised relative to the maximum intensity of that channel. The lines are Gaussian fits to the data, giving a mean beam waist radius of $0.67(6)\um$ and a mean channel spacing of $3.95(3)\um$. (a) A measurement of all 8 channels. (b) A zoomed-in view of the third channel [left-to-right in (a)]. In most cases the errorbars are smaller than the marker size.}
    \label{fig:waveguide-ion-data}
\end{figure}

The results from this measurement are shown in fig.~\ref{fig:waveguide-ion-data}. We measured a mean beam waist radius of $0.67(6)\um$ and a mean channel spacing of $3.95(3)\um$. For all channels, the cross-talk level was below $10^{-3}$, with a lowest measured cross-talk of $\mathcal{O}\left(10^{-5}\right)$. The variation in cross-talk across the chip was most likely due to light leakage through the fibre AOMs as well as aberrations and/or scatter from the optical components in the beam path that affect the channels non-uniformly.

\begin{figure}[!ht]
    \centering
    \includegraphics{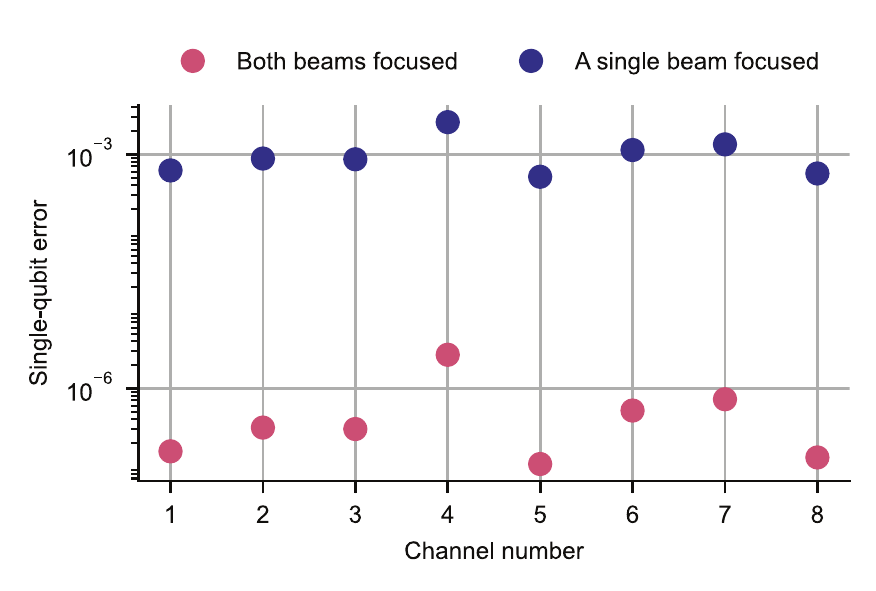}
    \caption{Estimated single-qubit error due to intensity cross-talk in the device accounting for contributions from all nearest-neighbour channels based on the measurements presented in fig.~\ref{fig:waveguide-ion-data}. The errorbars are smaller than the marker size.}
    \label{fig:single-qubit-error}
\end{figure}

In fig.~\ref{fig:single-qubit-error} we plot the error introduced to the state of a target ion's nearest-neighbours for the data shown in fig.~\ref{fig:waveguide-ion-data}. We consider two cases: when a single beam is focused down on the target ion, and the case of a Raman transition where both beams are focused on the target ion. The former is relevant when the qubit transition is itself an optical transition, or when a two-photon Raman process is used and one of the beams is global for the entire qubit register. For each channel we sum the nearest-neighbour contributions (except for the edge channels where there is a single contribution). In both cases this error is much lower than, or comparable to, state-of-the-art two-qubit gate errors\cite{ballance_high-fidelity_2016, wang_high-fidelity_2020, srinivas_high-fidelity_2021} and will therefore not limit the device performance. The cross-talk performance can be enhanced even further with well-known techniques such as coherent cancellation\cite{flannery_optical_2022} or composite pulse sequences\cite{brown_arbitrarily_2004}.

\subsection*{Discussions}
\label{sec:discussions}
We have presented a scalable and configurable way for individual optical addressing in chains of trapped atomic ions. We have designed and fabricated an SPIM-WG chip that offers a very high RI contrast between the core and the cladding, therefore achieving very low cross-talk at a small channel separation. The photonic chip is coupled with low loss to a fibre network that enables individual phase, frequency, and amplitude control of each of the channels, thus enabling parallel operations.

Measurements of the performance of our individual addressing system, optimised for $532\nm$ light, using a single trapped \ba ion showed cross-talk well below $10^{-3}$ across the chip. The corresponding estimated worst-case nearest-neighbour error is $\mathcal{O}\left(10^{-6}\right)$ if both beams are focused on a single ion or $\mathcal{O}\left(10^{-3}\right)$ if one beam is focused on the ion and the second beam is global for the register. In the latter case, well-known techniques such as composite pulses or coherent cancellation can be used to further reduce this error. 

The procedure used to manufacture the SPIM-WGs is highly flexible and so it can be used to create devices with a large number of channels, as well as devices with non-uniform channel spacing. The latter capability is important for applications in ion traps that do not have sufficient degrees of freedom to generate anharmonic potentials to keep the ion spacing uniform across the ion chain, or where harmonic potentials are desired. Preliminary results from the characterisation of devices with 32 channels show no degradation of performance in terms of cross-talk and minimal change in propagation losses. In addition, the same technique can be used to manufacture devices optimised for use at other optical wavelengths and therefore be integrated into setups using different ion species and/or different gate mechanisms.

In conclusion, we have presented a novel method to individually optically address chains of trapped atomic ions. Our method achieves significantly lower cross-talk compared to existing methods integrated with trapped-ion setups, while maintaining high scalability and flexibility.

\section*{Materials and methods}
\label{sec:materials}
\subsection*{Mode progression in the SPIM-WG chip}
\begin{figure}[!ht]
    \centering
    \includegraphics{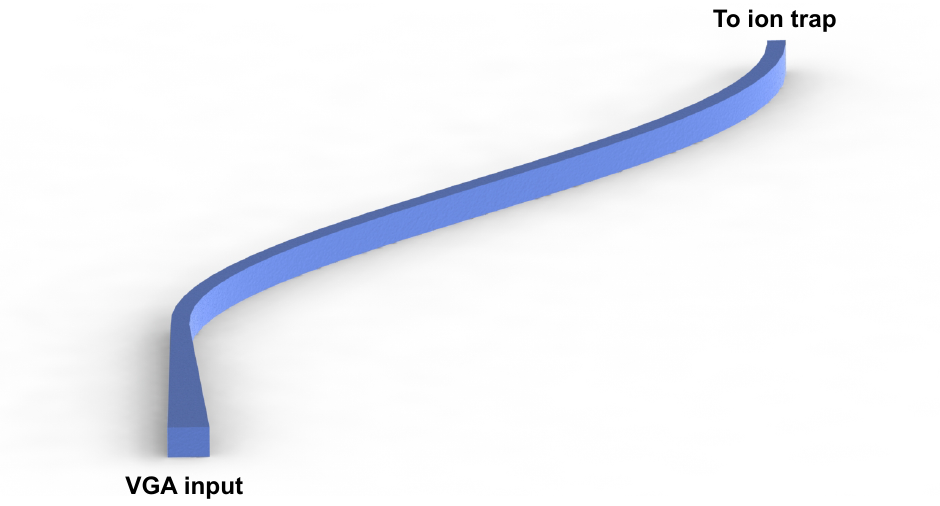}
    \caption{A 3D diagram of a single adiabatic mode converter channel in the photonic chip. The adiabatic mode conversion was implemented at the straight input region, while the channel cross-section at the bending region was kept constant until the channel output.}
    \label{fig:single-channel-3D}
\end{figure}

\subsection*{SPIM-WG chip integration into the ion trap setup}
\begin{figure}[!ht]
    \centering
    \includegraphics{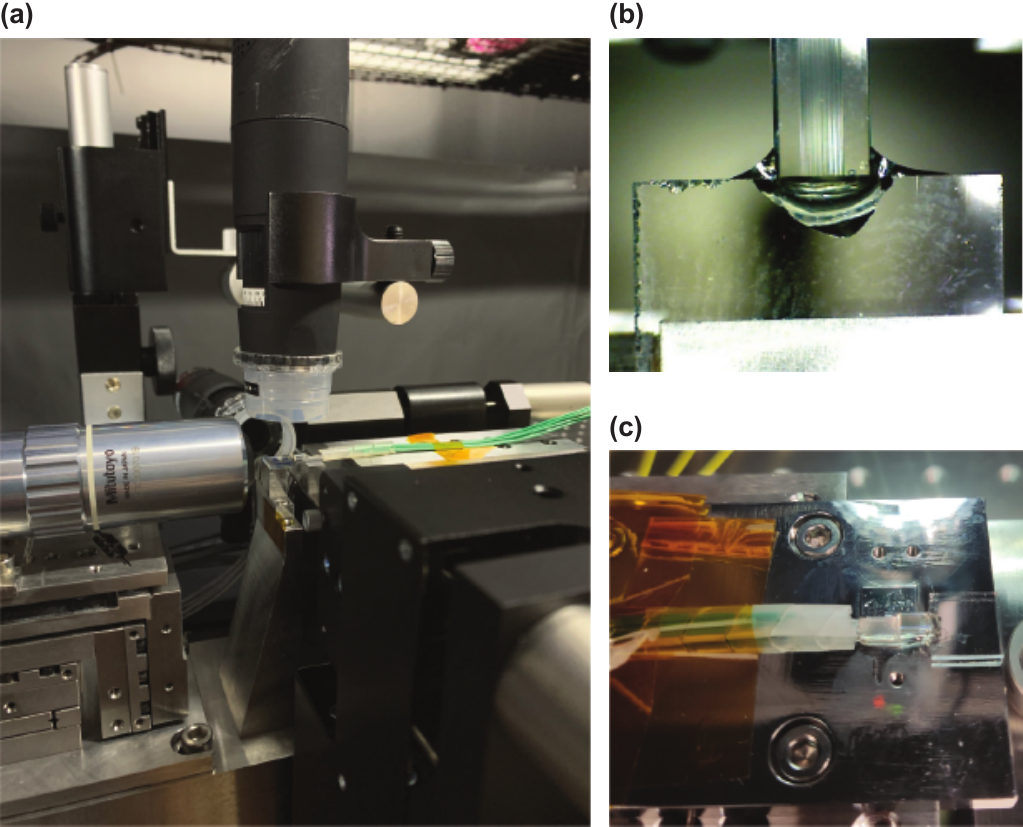}
    \caption{Coupling between the VGA and the photonic chip. (a) Setup used to couple light from the VGA into the photonic chip. The VGA was placed on a 6-axis positioning stage (right) while the photonic chip was held stationary (centre). The objective on the left was used to image the photonic chip output on a camera to evaluate the quality of the coupling. (b) A photo of the VGA glued to the photonic chip after the coupling was optimised. (c) A photo of the VGA and photonic chip assembly glued on a stainless steel plate prior to integration with the rest of the optical system in the ion trap setup.}
    \label{fig:waveguide-glue}
\end{figure}

To couple light from the VGA to the photonic chip, we kept the chip fixed and placed the VGA on a 6-axis positioning stage as shown in fig.~\ref{fig:waveguide-glue}. We imaged the chip output on a camera to evaluate the amount of light coupled from the VGA. We optimised the position of the VGA to maximise the average coupling efficiency, while simultaneously keeping the coupling efficiency across all channels as uniform as possible. Due to tolerances in the VGA spot size and spacing, the average coupling efficiency we achieved when optimising for all channels was lower than the maximum coupling efficiency that we could achieve for a single channel. In the devices used for the demonstration here, we achieved an average coupling efficiency of $45(3)\%$. 

As outlined in the main text, the coupling between the VGA and the chip is extremely sensitive to relative position changes between the VGA and the chip. In addition to changes in the power in each channel, changes in the coupling efficiency can also modify the phase and/or polarisation of the light that lead to gate errors. To ensure the coupling between the VGA and the chip stays constant during system operation, after optimising the coupling between the two components, the VGA was glued to the chip using the NOA061 UV curing glue as shown in fig.~\ref{fig:waveguide-glue}(b). To avoid changes in the coupling while the glue was curing, we cured at a gradually increasing UV power over $2-3$ hours.

The VGA and photonic chip assembly was then glued to a stainless steel plate and was transported to the ion trap system. This stainless steel plate was then bolted on a positioning stage that is part of the setup as shown in fig.~\ref{fig:waveguide-glue}(c). Upon exchanging chips, the stainless steel plate with the glued VGA-photonic chip assembly is the only component that needs to be exchanged, while the remaining optical setup is left unchanged. Therefore only minimal realignment is necessary following chip exchange. 

\subsection*{Robust phase estimation for an AC Stark shift measurement}
\begin{figure}[!ht]
    \centering
    \includegraphics
    {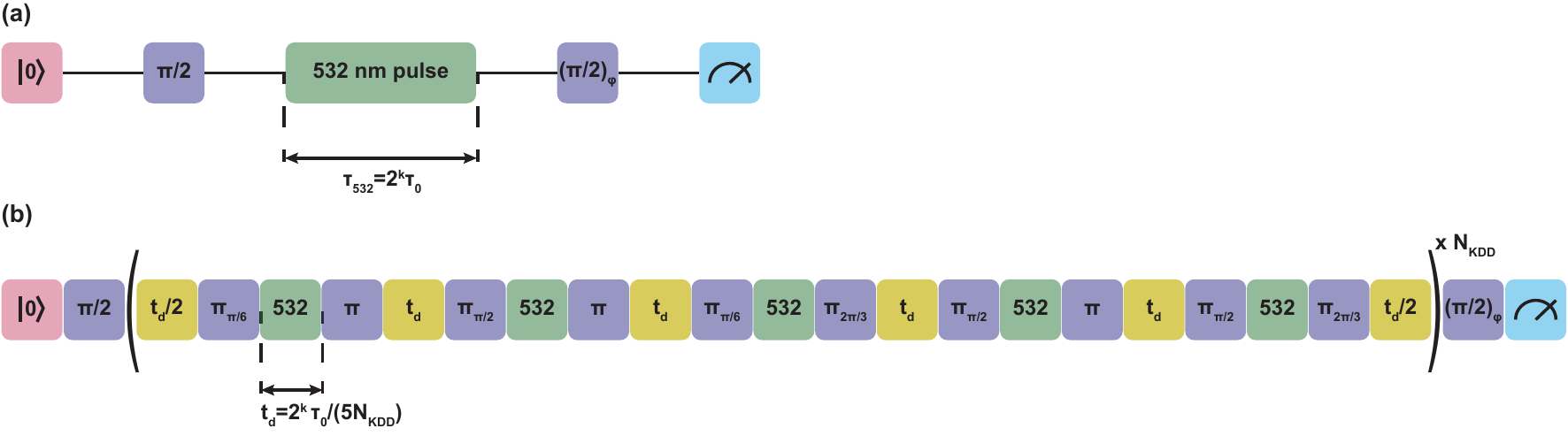}
    \caption{Pulse sequence for an AC Stark shift measurement. Unless otherwise specified, all $\pi$ and $\pi/2$ pulses are rotations about the $x$-axis of the Bloch sphere. (a) Robust phase estimation (RPE) sequence for estimating the phase introduced on the $\ket{\rightarrow}=\left(\ket{0}+i\ket{1}\right)/\sqrt{2}$ state as a result of the AC Stark shift from a single $532\nm$ beam. The ion is prepared in the state $\ket{0}$ and the subsequent $1762\nm$ $\pi/2$ pulse prepares the state $\ket{\rightarrow}$. Then a single $532\nm$ beam is turned on for a duration $\tau_{532}=2^k\tau_0$ where $\tau_0$ is the \textit{base} duration defining the maximum observable AC stark shift of $\delta_\mathrm{ac,\;max}=1/\tau_0$. The final $1762\nm$ $\left(\pi/2\right)_\phi$ pulse defines the measurement basis. The measurement is performed for $\phi=0, \pi/2, \pi, 3\pi/2$ to minimise the effect of polarising noise in one direction. (b) The sequence used to measure the AC Stark shift introduced by a single $532\nm$ beam on the $\ket{0}\leftrightarrow\ket{1}$ transition, consisting of the RPE protocol embedded into a KDD sequence to extend the system coherence time.}
    \label{fig:pulse-sequence}
\end{figure}

We used the AC Stark shift induced by a single $532\nm$ beam on the $\ket{S_{1/2}, F=2, m_F=0}\leftrightarrow\ket{D_{5/2}, F=4, m_F=+1}$ transition to measure the intensity of the beams and hence characterise the chip output at the ion position. This AC Stark shift introduces a relative phase between the two states that is proportional to the beam intensity. We used robust phase estimation \cite{kimmel_robust_2015, rudinger_experimental_2017} (RPE) as shown in fig.~\ref{fig:pulse-sequence}(a) to estimate that phase. This significantly reduced our sensitivity to other systematic errors such as state preparation and measurement errors. To reduce the sensitivity of the system to decoherence and enable longer probing durations for the RPE sequence (hence also increasing the measurement dynamic range) we embedded the RPE sequence as part of a Knill dynamical decoupling sequence\cite{souza_robust_2011} (KDD) as shown in fig.~\ref{fig:pulse-sequence}(b). For probe durations shorter than the system coherence time, the RPE sequence was embedded in a spin-echo sequence instead. This avoided $532\nm$ pulses with durations comparable to the AOM settling time constants. For probe durations longer than the system coherence time, the KDD sequence in fig.~\ref{fig:pulse-sequence}(b) was used and the number of KDD pulses $N_\mathrm{KDD}$ was tuned to ensure that the maximum duration between consecutive $\pi$-pulses did not exceed the system coherence time. In this experiment this duration was set to $500\us$.

\section*{Acknowledgements}
We thank A. Sinclair, G. Wilpers, and team at NPL for providing the ion trap and vacuum package used in this work. We thank Viviene Dela Cruz for contributing to the SPIM-WG characterisation. This work was supported by a UKRI FL Fellowship (MR/S03238X/1); the US Army Research Office (W911NF-20-1-0038); the UK EPSRC Hub in Quantum Computing and Simulation (EP/T001062/1); EPSRC Fellowship (EP/T00326X/1); Marie Curie Fellowship UKRI guarantee (EP/X024296/1); Austrian Science Fund (I3984-N36). C.H. acknowledges St. John’s College, Oxford for support through a Junior Research Fellowship. D.P.N. thanks Merton College, Oxford for the same. A.S.S. acknowledges funding from the JT Hamilton scholarship from Balliol College, Oxford.

\section*{Author contributions}

B.S. designed and fabricated the SPIM-WG photonic chips. B.S. conducted various simulations and performed the classical tests of the chips. A.S.S. designed and implemented the systems and procedures used to integrate the chips with the ion trap apparatus. A.S.S. and D.P.N. took the single-ion measurements. A.S.S. performed the data analysis. M.W. and A.W. assisted with the photonic chip fabrication, characterization and waveguide mode analysis. C.H. analysed the polarisation and phase effects of the waveguide coupling errors, and assisted with microscopy. A.J. and S.M. performed the measurements and the analysis of the waveguide refractive index properties. A.S.S. and F.P. set up the ion trap experiment. J.D.L., A.V.B., and F.P. contributed to the experiment apparatus and supported the operation of the experiment. M.J.B.and C.J.B. obtained funding and supervised the project. A.S.S. and B.S. wrote the manuscript, with assistance from M.J.B and C.J.B. A.W. assisted with figure preparation. All authors reviewed the manuscript. 

\section*{Conflict of interest}
The authors declare no competing interests.

\bibliography{main}

\begin{thebibliography}{10}
\urlstyle{rm}
\expandafter\ifx\csname url\endcsname\relax
  \def\url#1{\texttt{#1}}\fi
\expandafter\ifx\csname urlprefix\endcsname\relax\def\urlprefix{URL }\fi
\expandafter\ifx\csname doiprefix\endcsname\relax\def\doiprefix{DOI: }\fi
\providecommand{\bibinfo}[2]{#2}
\providecommand{\eprint}[2][]{\url{#2}}

\bibitem{cirac_quantum_1995}
\bibinfo{author}{Cirac, J.~I.} \& \bibinfo{author}{Zoller, P.}
\newblock \bibinfo{journal}{\bibinfo{title}{Quantum {Computations} with {Cold}
  {Trapped} {Ions}}}.
\newblock {\emph{\JournalTitle{Physical Review Letters}}}
  \textbf{\bibinfo{volume}{74}}, \bibinfo{pages}{4091--4094}
  (\bibinfo{year}{1995}).

\bibitem{harty_high-fidelity_2014}
\bibinfo{author}{Harty, T.} \emph{et~al.}
\newblock \bibinfo{journal}{\bibinfo{title}{High-{Fidelity} {Preparation},
  {Gates}, {Memory}, and {Readout} of a {Trapped}-{Ion} {Quantum} {Bit}}}.
\newblock {\emph{\JournalTitle{Physical Review Letters}}}
  \textbf{\bibinfo{volume}{113}}, \bibinfo{pages}{220501}
  (\bibinfo{year}{2014}).

\bibitem{ballance_high-fidelity_2016}
\bibinfo{author}{Ballance, C.}, \bibinfo{author}{Harty, T.},
  \bibinfo{author}{Linke, N.}, \bibinfo{author}{Sepiol, M.} \&
  \bibinfo{author}{Lucas, D.}
\newblock \bibinfo{journal}{\bibinfo{title}{High-{Fidelity} {Quantum} {Logic}
  {Gates} {Using} {Trapped}-{Ion} {Hyperfine} {Qubits}}}.
\newblock {\emph{\JournalTitle{Physical Review Letters}}}
  \textbf{\bibinfo{volume}{117}}, \bibinfo{pages}{060504}
  (\bibinfo{year}{2016}).

\bibitem{clark_high-fidelity_2021}
\bibinfo{author}{Clark, C.~R.} \emph{et~al.}
\newblock \bibinfo{journal}{\bibinfo{title}{High-{Fidelity} {Bell}-{State}
  {Preparation} with $^{40}\mathrm{Ca}^+$ {Optical} {Qubits}}}.
\newblock {\emph{\JournalTitle{Physical Review Letters}}}
  \textbf{\bibinfo{volume}{127}}, \bibinfo{pages}{130505}
  (\bibinfo{year}{2021}).

\bibitem{srinivas_high-fidelity_2021}
\bibinfo{author}{Srinivas, R.} \emph{et~al.}
\newblock \bibinfo{journal}{\bibinfo{title}{High-fidelity laser-free universal
  control of trapped ion qubits}}.
\newblock {\emph{\JournalTitle{Nature}}} \textbf{\bibinfo{volume}{597}},
  \bibinfo{pages}{209--213} (\bibinfo{year}{2021}).

\bibitem{wang_single_2021}
\bibinfo{author}{Wang, P.} \emph{et~al.}
\newblock \bibinfo{journal}{\bibinfo{title}{Single ion qubit with estimated
  coherence time exceeding one hour}}.
\newblock {\emph{\JournalTitle{Nature Communications}}}
  \textbf{\bibinfo{volume}{12}}, \bibinfo{pages}{233} (\bibinfo{year}{2021}).

\bibitem{an_high_2022}
\bibinfo{author}{An, F.~A.} \emph{et~al.}
\newblock \bibinfo{journal}{\bibinfo{title}{High {Fidelity} {State}
  {Preparation} and {Measurement} of {Ion} {Hyperfine} {Qubits} with
  ${I}>1/2$}}.
\newblock {\emph{\JournalTitle{Physical Review Letters}}}
  \textbf{\bibinfo{volume}{129}}, \bibinfo{pages}{130501}
  (\bibinfo{year}{2022}).

\bibitem{wineland_experimental_1998}
\bibinfo{author}{Wineland, D.} \emph{et~al.}
\newblock \bibinfo{journal}{\bibinfo{title}{Experimental issues in coherent
  quantum-state manipulation of trapped atomic ions}}.
\newblock {\emph{\JournalTitle{Journal of Research of the National Institute of
  Standards and Technology}}} \textbf{\bibinfo{volume}{103}},
  \bibinfo{pages}{259} (\bibinfo{year}{1998}).

\bibitem{nagerl_ion_1998}
\bibinfo{author}{Nägerl, H.}, \bibinfo{author}{Bechter, W.},
  \bibinfo{author}{Eschner, J.}, \bibinfo{author}{Schmidt-Kaler, F.} \&
  \bibinfo{author}{Blatt, R.}
\newblock \bibinfo{journal}{\bibinfo{title}{Ion strings for quantum gates}}.
\newblock {\emph{\JournalTitle{Applied Physics B}}}
  \textbf{\bibinfo{volume}{66}}, \bibinfo{pages}{603--608}
  (\bibinfo{year}{1998}).

\bibitem{leu_fast_2023}
\bibinfo{author}{Leu, A.~D.} \emph{et~al.}
\newblock \bibinfo{title}{Fast, high-fidelity addressed single-qubit gates
  using efficient composite pulse sequences} (\bibinfo{year}{2023}).
\newblock \bibinfo{note}{ArXiv:2305.06725 [quant-ph]}.

\bibitem{srinivas_coherent_2023}
\bibinfo{author}{Srinivas, R.} \emph{et~al.}
\newblock \bibinfo{journal}{\bibinfo{title}{Coherent {Control} of
  {Trapped}-{Ion} {Qubits} with {Localized} {Electric} {Fields}}}.
\newblock {\emph{\JournalTitle{Physical Review Letters}}}
  \textbf{\bibinfo{volume}{131}}, \bibinfo{pages}{020601}
  (\bibinfo{year}{2023}).

\bibitem{leibfried_individual_1999}
\bibinfo{author}{Leibfried, D.}
\newblock \bibinfo{journal}{\bibinfo{title}{Individual addressing and state
  readout of trapped ions utilizing rf micromotion}}.
\newblock {\emph{\JournalTitle{Physical Review A}}}
  \textbf{\bibinfo{volume}{60}}, \bibinfo{pages}{R3335--R3338}
  (\bibinfo{year}{1999}).

\bibitem{warring_individual-ion_2013}
\bibinfo{author}{Warring, U.} \emph{et~al.}
\newblock \bibinfo{journal}{\bibinfo{title}{Individual-{Ion} {Addressing} with
  {Microwave} {Field} {Gradients}}}.
\newblock {\emph{\JournalTitle{Physical Review Letters}}}
  \textbf{\bibinfo{volume}{110}}, \bibinfo{pages}{173002}
  (\bibinfo{year}{2013}).

\bibitem{wang_individual_2009}
\bibinfo{author}{Wang, S.~X.}, \bibinfo{author}{Labaziewicz, J.},
  \bibinfo{author}{Ge, Y.}, \bibinfo{author}{Shewmon, R.} \&
  \bibinfo{author}{Chuang, I.~L.}
\newblock \bibinfo{journal}{\bibinfo{title}{Individual addressing of ions using
  magnetic field gradients in a surface-electrode ion trap}}.
\newblock {\emph{\JournalTitle{Applied Physics Letters}}}
  \textbf{\bibinfo{volume}{94}}, \bibinfo{pages}{094103}
  (\bibinfo{year}{2009}).

\bibitem{seck_single-ion_2020}
\bibinfo{author}{Seck, C.~M.} \emph{et~al.}
\newblock \bibinfo{journal}{\bibinfo{title}{Single-ion addressing via trap
  potential modulation in global optical fields}}.
\newblock {\emph{\JournalTitle{New Journal of Physics}}}
  \textbf{\bibinfo{volume}{22}}, \bibinfo{pages}{053024}
  (\bibinfo{year}{2020}).

\bibitem{crain_individual_2014}
\bibinfo{author}{Crain, S.}, \bibinfo{author}{Mount, E.},
  \bibinfo{author}{Baek, S.} \& \bibinfo{author}{Kim, J.}
\newblock \bibinfo{journal}{\bibinfo{title}{Individual addressing of trapped
  $^{171}\mathrm{Yb}^+$ ion qubits using a microelectromechanical systems-based
  beam steering system}}.
\newblock {\emph{\JournalTitle{Applied Physics Letters}}}
  \textbf{\bibinfo{volume}{105}}, \bibinfo{pages}{181115}
  (\bibinfo{year}{2014}).

\bibitem{shih_reprogrammable_2021}
\bibinfo{author}{Shih, C.-Y.} \emph{et~al.}
\newblock \bibinfo{journal}{\bibinfo{title}{Reprogrammable and high-precision
  holographic optical addressing of trapped ions for scalable quantum
  control}}.
\newblock {\emph{\JournalTitle{npj Quantum Information}}}
  \textbf{\bibinfo{volume}{7}}, \bibinfo{pages}{1--8} (\bibinfo{year}{2021}).

\bibitem{wang_high-fidelity_2020}
\bibinfo{author}{Wang, Y.} \emph{et~al.}
\newblock \bibinfo{journal}{\bibinfo{title}{High-{Fidelity} {Two}-{Qubit}
  {Gates} {Using} a {Microelectromechanical}-{System}-{Based} {Beam} {Steering}
  {System} for {Individual} {Qubit} {Addressing}}}.
\newblock {\emph{\JournalTitle{Physical Review Letters}}}
  \textbf{\bibinfo{volume}{125}}, \bibinfo{pages}{150505}
  (\bibinfo{year}{2020}).

\bibitem{pogorelov_compact_2021}
\bibinfo{author}{Pogorelov, I.} \emph{et~al.}
\newblock \bibinfo{journal}{\bibinfo{title}{Compact {Ion}-{Trap} {Quantum}
  {Computing} {Demonstrator}}}.
\newblock {\emph{\JournalTitle{PRX Quantum}}} \textbf{\bibinfo{volume}{2}},
  \bibinfo{pages}{020343} (\bibinfo{year}{2021}).

\bibitem{egan_scaling_2021}
\bibinfo{author}{Egan, L.~N.}
\newblock \emph{\bibinfo{title}{{SCALING} {QUANTUM} {COMPUTERS} {WITH} {LONG}
  {CHAINS} {OF} {TRAPPED} {IONS}}}.
\newblock Ph.D. thesis, \bibinfo{school}{Unviersity of Maryland}
  (\bibinfo{year}{2021}).

\bibitem{binai-motlagh_guided_2023}
\bibinfo{author}{Binai-Motlagh, A.} \emph{et~al.}
\newblock \bibinfo{title}{A guided light system for agile individual addressing
  of $\mathrm{Ba}^+$ qubits with $10^{-4}$ level intensity crosstalk}
  (\bibinfo{year}{2023}).
\newblock \bibinfo{note}{ArXiv:2302.14711 [physics, physics:quant-ph]}.

\bibitem{timpu_laser-written_2022}
\bibinfo{author}{Timpu, F.} \emph{et~al.}
\newblock \bibinfo{title}{Laser-{Written} {Waveguide} {Array} {Optimized} for
  {Individual} {Control} of {Trapped} {Ion} {Qubits} in a {Chain}}.
\newblock In \emph{\bibinfo{booktitle}{2022 {European} {Conference} on
  {Optical} {Communication} ({ECOC})}}, \bibinfo{pages}{1--4}
  (\bibinfo{year}{2022}).

\bibitem{wilpers_compact_2013}
\bibinfo{author}{Wilpers, G.}, \bibinfo{author}{See, P.},
  \bibinfo{author}{Gill, P.} \& \bibinfo{author}{Sinclair, A.~G.}
\newblock \bibinfo{journal}{\bibinfo{title}{A compact {UHV} package for
  microfabricated ion-trap arrays with direct electronic air-side access}}.
\newblock {\emph{\JournalTitle{Applied Physics B}}}
  \textbf{\bibinfo{volume}{111}}, \bibinfo{pages}{21--28}
  (\bibinfo{year}{2013}).

\bibitem{choonee_silicon_2017}
\bibinfo{author}{Choonee, K.}, \bibinfo{author}{Wilpers, G.} \&
  \bibinfo{author}{Sinclair, A.~G.}
\newblock \bibinfo{title}{Silicon microfabricated linear segmented ion traps
  for quantum technologies}.
\newblock In \emph{\bibinfo{booktitle}{2017 19th {International} {Conference}
  on {Solid}-{State} {Sensors}, {Actuators} and {Microsystems}
  ({TRANSDUCERS})}}, \bibinfo{pages}{615--618} (\bibinfo{year}{2017}).

\bibitem{sun_-chip_2022}
\bibinfo{author}{Sun, B.} \emph{et~al.}
\newblock \bibinfo{journal}{\bibinfo{title}{On-chip beam rotators, adiabatic
  mode converters, and waveplates through low-loss waveguides with variable
  cross-sections}}.
\newblock {\emph{\JournalTitle{Light: Science \& Applications}}}
  \textbf{\bibinfo{volume}{11}}, \bibinfo{pages}{214} (\bibinfo{year}{2022}).

\bibitem{cetina_control_2022}
\bibinfo{author}{Cetina, M.} \emph{et~al.}
\newblock \bibinfo{journal}{\bibinfo{title}{Control of {Transverse} {Motion}
  for {Quantum} {Gates} on {Individually} {Addressed} {Atomic} {Qubits}}}.
\newblock {\emph{\JournalTitle{PRX Quantum}}} \textbf{\bibinfo{volume}{3}},
  \bibinfo{pages}{010334} (\bibinfo{year}{2022}).

\bibitem{west_tunable_2021}
\bibinfo{author}{West, A.~D.}, \bibinfo{author}{Putnam, R.},
  \bibinfo{author}{Campbell, W.~C.} \& \bibinfo{author}{Hamilton, P.}
\newblock \bibinfo{journal}{\bibinfo{title}{Tunable transverse spin–motion
  coupling for quantum information processing}}.
\newblock {\emph{\JournalTitle{Quantum Science and Technology}}}
  \textbf{\bibinfo{volume}{6}}, \bibinfo{pages}{024003} (\bibinfo{year}{2021}).

\bibitem{sun_high_2023}
\bibinfo{author}{Sun, B.} \emph{et~al.}
\newblock \bibinfo{title}{High {Speed} {Precise} {Refractive} {Index}
  {Modification} for {Photonic} {Chips} through {Phase} {Aberrated} {Pulsed}
  {Lasers}} (\bibinfo{year}{2023}).
\newblock \bibinfo{note}{ArXiv:2307.14451 [physics]}.

\bibitem{snyder_optical_1984}
\bibinfo{author}{Snyder, A.~W.} \& \bibinfo{author}{Love, J.~D.}
\newblock \emph{\bibinfo{title}{Optical {Waveguide} {Theory}}}
  (\bibinfo{address}{Boston, MA}, \bibinfo{year}{1984}).

\bibitem{lin_large-scale_2009}
\bibinfo{author}{Lin, G.-D.} \emph{et~al.}
\newblock \bibinfo{journal}{\bibinfo{title}{Large-scale quantum computation in
  an anharmonic linear ion trap}}.
\newblock {\emph{\JournalTitle{Europhysics Letters}}}
  \textbf{\bibinfo{volume}{86}}, \bibinfo{pages}{60004} (\bibinfo{year}{2009}).

\bibitem{zhang_branching_2020}
\bibinfo{author}{Zhang, Z.} \emph{et~al.}
\newblock \bibinfo{journal}{\bibinfo{title}{Branching fractions for
  $\mathrm{P}_{3/2}$ decays in $\mathrm{Ba}^+$}}.
\newblock {\emph{\JournalTitle{Physical Review A}}}
  \textbf{\bibinfo{volume}{101}}, \bibinfo{pages}{062515}
  (\bibinfo{year}{2020}).

\bibitem{kimmel_robust_2015}
\bibinfo{author}{Kimmel, S.}, \bibinfo{author}{Low, G.~H.} \&
  \bibinfo{author}{Yoder, T.~J.}
\newblock \bibinfo{journal}{\bibinfo{title}{Robust calibration of a universal
  single-qubit gate set via robust phase estimation}}.
\newblock {\emph{\JournalTitle{Physical Review A}}}
  \textbf{\bibinfo{volume}{92}}, \bibinfo{pages}{062315}
  (\bibinfo{year}{2015}).

\bibitem{rudinger_experimental_2017}
\bibinfo{author}{Rudinger, K.}, \bibinfo{author}{Kimmel, S.},
  \bibinfo{author}{Lobser, D.} \& \bibinfo{author}{Maunz, P.}
\newblock \bibinfo{journal}{\bibinfo{title}{Experimental {Demonstration} of a
  {Cheap} and {Accurate} {Phase} {Estimation}}}.
\newblock {\emph{\JournalTitle{Physical Review Letters}}}
  \textbf{\bibinfo{volume}{118}}, \bibinfo{pages}{190502}
  (\bibinfo{year}{2017}).

\bibitem{souza_robust_2011}
\bibinfo{author}{Souza, A.~M.}, \bibinfo{author}{Álvarez, G.~A.} \&
  \bibinfo{author}{Suter, D.}
\newblock \bibinfo{journal}{\bibinfo{title}{Robust {Dynamical} {Decoupling} for
  {Quantum} {Computing} and {Quantum} {Memory}}}.
\newblock {\emph{\JournalTitle{Physical Review Letters}}}
  \textbf{\bibinfo{volume}{106}}, \bibinfo{pages}{240501}
  (\bibinfo{year}{2011}).

\bibitem{flannery_optical_2022}
\bibinfo{author}{Flannery, J.} \emph{et~al.}
\newblock \bibinfo{title}{Optical {Crosstalk} {Mitigation} for {Individual}
  {Addressing} in a {Cryogenic} {Ion} {Trap}}.
\newblock In \emph{\bibinfo{booktitle}{2022 {IEEE} {International} {Conference}
  on {Quantum} {Computing} and {Engineering} ({QCE})}},
  \bibinfo{pages}{816--817} (\bibinfo{year}{2022}).

\bibitem{brown_arbitrarily_2004}
\bibinfo{author}{Brown, K.~R.}, \bibinfo{author}{Harrow, A.~W.} \&
  \bibinfo{author}{Chuang, I.~L.}
\newblock \bibinfo{journal}{\bibinfo{title}{Arbitrarily accurate composite
  pulse sequences}}.
\newblock {\emph{\JournalTitle{Physical Review A}}}
  \textbf{\bibinfo{volume}{70}}, \bibinfo{pages}{052318}
  (\bibinfo{year}{2004}).

\end{thebibliography}

\end{document}